\documentclass[aps,prd,reprint,nofootinbib]{revtex4-2}
\usepackage[utf8]{inputenc}

\usepackage{mathtools}
\usepackage{physics}
\usepackage{booktabs}
\usepackage{array}
\usepackage{xcolor}
\usepackage{bm}
\usepackage[hidelinks]{hyperref}

\newcommand{\PreserveBackslash}[1]{\let\temp=\\#1\let\\=\temp}
\newcolumntype{C}[1]{>{\PreserveBackslash\centering}p{#1}}
\newcolumntype{R}[1]{>{\PreserveBackslash\raggedleft}p{#1}}
\newcolumntype{L}[1]{>{\PreserveBackslash\raggedright}p{#1}}

\newcommand{\overbar}[1]{\mkern 3.5mu\overline{\mkern-3.5mu#1}}
\newcommand{\xioverbar}[1]{\mkern 2mu\overline{\mkern-2mu#1}}

\newcommand{\R}{\mathcal{R}}
\newcommand{\Pp}{\mathcal{P}}
\newcommand{\PR}{\mathcal{P}_\mathcal{R}}
\newcommand{\Pphi}{\mathcal{P}_\phi}
\newcommand{\Ppi}{\mathcal{P}_\pi}
\newcommand{\Pphipi}{\mathcal{P}_{\phi\pi}}
\newcommand{\kc}{k_\sigma}
\newcommand{\kini}{k_\text{ini}}
\newcommand{\dN}{\Delta N}
\newcommand{\bdN}{\overbar{\dN}}
\newcommand{\eps}{\epsilon}
\newcommand{\epseff}{\tilde{\epsilon}}
\newcommand{\phieff}{\tilde{\phi}}
\newcommand{\pieff}{\tilde{\pi}}
\newcommand{\dphi}{\delta\phi}
\newcommand{\Dphi}{\Delta \phi}
\newcommand{\Neff}{\tilde{N}}

\newcommand{\xiN}{\hat{\xi}_N}
\newcommand{\xiNd}{\hat{\xi}_{N'}}
\newcommand{\xik}{\hat{\xi}_{k}}
\newcommand{\xikd}{\hat{\xi}_{k'}}

\newcommand{\sig}[1]{\sigma_{#1}}

\newcommand{\sigC}[2]{\sigma_{#1|#2}}

\newcommand{\erfc}{\text{erfc}}
\newcommand{\Ito}{It\^{o}}

\begin{document}

\title{Stochastic constant-roll inflation and primordial black holes}
\author{Eemeli Tomberg}
\email{eemeli.tomberg@kbfi.ee}
\affiliation{Laboratory of High Energy and Computational Physics, National Institute
of Chemical Physics and Biophysics, R\"{a}vala pst. 10, 10143 Tallinn, Estonia}
\date{\today}

\keywords{cosmology, early universe, stochastic inflation, constant roll, primordial black holes}

\begin{abstract} Stochastic inflation resolves primordial perturbations non-linearly, probing their probability distribution deep into its non-Gaussian tail. The strongest perturbations collapse into primordial black holes. In typical black-hole-producing single-field inflation, the strongest stochastic kicks occur during a period of constant roll. In this paper, I solve the stochastic constant-roll system, drawing the stochastic kicks from a numerically computed power spectrum, beyond the usual de Sitter approximation. The perturbation probability distribution is an analytical function of the integrated curvature power spectrum $\sig{k}^2$ and the second slow-roll parameter $\eps_2$. With a large $\eps_2$, stochastic effects can reduce the height of the curvature power spectrum required to form asteroid mass black holes from $10^{-2}$ to $10^{-3}$. I compare these results to studies with the non-stochastic $\Delta N$ formalism.
\end{abstract}

\maketitle

\section{Introduction}
During cosmic inflation, quantum vacuum fluctuations stretch and grow, forming the seeds of the classical structure of the late universe. Studying these fluctuations provides a glimpse into the quantum nature of gravity and offers insight into the interplay between quantum and classical physics. This interplay is made manifest in \emph{stochastic inflation} \cite{Starobinsky:1986fx}, where a classically evolving coarse-grained FLRW (Friedmann--Lema\^itre--Robertson--Walker) universe receives random kicks from short-wavelength quantum fluctuations.

Combined with the $\dN$ formalism \cite{Sasaki:1995aw, Sasaki:1998ug, Wands:2000dp, Lyth:2004gb}, stochastic inflation describes the super-Hubble fluctuations non-perturbatively \cite{Fujita:2013cna, Vennin:2015hra}. This is important for primordial black holes (PBHs) \cite{Carr:1974nx, Carr:1975qj}, a dark matter candidate \cite{Chapline:1975ojl, Green:2020jor, Carr:2021bzv} forming from the rarest, strongest perturbations. Many recent works \cite{Pattison:2017mbe, Cruces:2018cvq, Ezquiaga:2018gbw, Biagetti:2018pjj, Firouzjahi:2018vet, Ezquiaga:2019ftu, Pattison:2019hef, Prokopec:2019srf, Ando:2020fjm, De:2020hdo, Figueroa:2020jkf, Ballesteros:2020sre, Cruces:2021iwq, Rigopoulos:2021nhv, Pattison:2021oen, Achucarro:2021pdh, Hooshangi:2021ubn, Tomberg:2021xxv, Figueroa:2021zah, Tada:2021zzj, Cruces:2022imf, Ahmadi:2022lsm, Animali:2022otk, Jackson:2022unc, Rigopoulos:2022gso, Briaud:2023eae, Mishra:2023lhe, Asadi:2023flu} study the stochastic system, employing simplifying assumptions to make the computation manageable or relying on heavy-duty numerics. They show that the stochastic effects enhance PBH abundance over the usual Gaussian approximation.

The black-hole-forming inflationary models typically contain a phase of ultra-slow-roll inflation, where the inflaton field climbs up towards a local maximum in its potential and its perturbations grow (see e.g. \cite{Kannike:2017bxn, Bezrukov:2017dyv, Ballesteros:2017fsr, Rasanen:2018fom}), followed by a dual phase of constant-roll inflation when the field rolls down the other side. The importance of the constant-roll phase was recently emphasized in \cite{Karam:2022nym}. In \cite{Tomberg:2022mkt}, it was pointed out that the coarse-grained field experiences the strongest stochastic motion precisely during constant roll, when the perturbations amplified by ultra slow roll provide their kicks. It was also shown that in such a setup, the inflaton's stochastic motion is constrained on its classical trajectory and that the stochastic kicks can be drawn from a pre-computed power spectrum $\PR(k)$, computed beyond the de Sitter approximation. This improves the computation's accuracy compared with previous studies.

In this paper, I refine the analysis of \cite{Tomberg:2022mkt}, solving the stochastic equations exactly in the pure constant-roll limit. I present an analytical expression for the probability distribution of the perturbations, interpolating between a central Gaussian and an exponential tail, and study the stochastic trajectories corresponding to a given perturbation strength. Finally, I discuss the consequences for PBH formation and compare the results to previous studies.

\section{Stochastic inflation}
I study an inflaton field $\varphi$ with the canonical action
\begin{equation} \label{eq:S}
    S = \int \dd^4 x \sqrt{-g}\qty[\frac{1}{2} R - \frac{1}{2}\partial^\mu\varphi\partial_\mu\varphi - V(\varphi)] \, ,
\end{equation}
where $g$ is the metric determinant, $R$ is the curvature scalar, $V$ is the inflaton potential, and the reduced Planck mass has been set to one. In the stochastic formalism, we divide the field into long and short wavelength parts, $\varphi = \phi + \dphi$, separated by the coarse-graining scale with Fourier wavenumber $\kc$:
\begin{equation} \label{eq:phi_division}
\begin{aligned}
    \phi &\equiv
    \int_{k<\kc} \frac{\dd^3 k}{(2\pi)^{2/3}}\varphi_k(N) e^{-i\vec{k} \cdot \vec{x}} \, , \\
    \delta \phi &\equiv
    \int_{k>\kc} \frac{\dd^3 k}{(2\pi)^{2/3}}\varphi_k(N) e^{-i\vec{k} \cdot \vec{x}} \, , \\
    \kc &\equiv \sigma aH_0 \, .
\end{aligned}
\end{equation}
In each super-Hubble patch, the long-wavelength part behaves like a local FLRW universe. Above, $a$ is the scale factor of this universe, $H$ is the Hubble parameter and $H_0$ its initial value\footnote{\label{ft:k}Some previous studies used $\kc = \sigma a H$ (see e.g. \cite{Pattison:2019hef, Figueroa:2021zah, Tomberg:2022mkt}). I use a constant $H_0$ instead of a time-dependent $H$ so that $\kc$ is a function of the number of e-folds $N=\ln a$ only. This makes the treatment simpler and more consistent and removes a factor of $(1-\eps_1)$ from the noise correlators \eqref{eq:noise_correlators}. Since $H$ changes little during inflation, the practical effect is small.}, and $\sigma \ll 1$ is a constant.

The long-wavelength part is classical, while the short-wavelength perturbations are quantum. If we neglect the latter, the former would follow the standard Friedmann equations
\begin{equation} \label{eq:bg_homog}
    \ddot{\phi} + 3H\dot{\phi} + V'(\phi) = 0 \, , \quad 3H^2 = \frac{1}{2}\dot{\phi}^2 + V(\phi) \, ,
\end{equation}
where dot denotes a derivative with respect to the cosmic time. However, the evolution of $\phi$ is also affected by the short-wavelength modes through the time-dependent $\kc$: the short modes are stretched by the expansion of the universe, and when they outgrow the coarse-graining scale, they join the background, changing it in a random quantum kick. The resulting stochastic equations for $\phi$ read (see e.g. \cite{Pattison:2019hef, Figueroa:2021zah, Tomberg:2022mkt})
\begin{equation} \label{eq:bg_eom}
    \phi' = \pi + \xi_\phi \, ,
    \quad
    \pi' = -\qty(3 - \eps_1)\qty(\pi + \frac{V_{,\phi}(\phi)}{V(\phi)}) + \xi_\pi \, ,
\end{equation}
where we have eliminated $H$, separated the second-order equation in \eqref{eq:bg_homog} into two first-order equations by introducing the field momentum $\pi$, and adopted the number of e-folds of expansion $N\equiv\ln a$ as the new time variable. Prime denotes a derivative with respect to $N$, and $\eps_1\equiv\frac{1}{2}\pi^2$ is the first slow-roll parameter. The $\xi_\phi$ and $\xi_\pi$ terms provide the stochastic kicks.

The short-wavelength perturbations live in the local background, and we evolve them linearly, Fourier mode by Fourier mode, using the Sasaki--Mukhanov equation. Due to linearity, the perturbations are Gaussian. The mode that is exiting the coarse-graining scale gives the two-point correlators of the stochastic kicks,
\begin{equation} \label{eq:noise_correlators}
\begin{aligned}
    \expval{\xi_\phi(N)\xi_\phi(N')} &=
    \Pphi(N,\kc)\delta(N-N') \, , \\
    \expval{\xi_\pi(N)\xi_\pi(N')} &=
    \Ppi(N,\kc)\delta(N-N') \, , \\
    \expval{\xi_\phi(N)\xi_\pi(N')} &=
    \Pphipi(N,\kc)\delta(N-N') \, ,
\end{aligned}
\end{equation}
where $\Pp_{X\!X}$ are the power spectra of the field and momentum perturbations in the spatially flat gauge,
\begin{equation} \label{eq:power_spectra}
\begin{aligned}
    \Pphi(N,k) &\equiv \frac{k^3}{2\pi^2}|\dphi_k(N)|^2 \, , \\
    \Ppi(N,k) &\equiv \frac{k^3}{2\pi^2}|\dphi'_k(N)|^2 \, , \\
    \Pphipi(N,k) &\equiv \frac{k^3}{2\pi^2}\dphi_k(N)\dphi'^*_k(N) \, .
\end{aligned}
\end{equation}
Despite appearances, there is only one independent kick at each time step: the quantum state of the short-wavelength perturbations is highly squeezed, leading to the strong correlation $\xi_\pi = \frac{\delta\phi'_k}{\delta\phi_k}\xi_\phi$ \cite{Figueroa:2020jkf}. Moreover, the kicks are aligned in a very specific way. To see this, consider the ratio of the comoving curvature perturbation $\R_k = \delta\phi_k/\pi$ and its derivative in the non-stochastic limit:
\begin{equation} \label{eq:Rprime_over_R}
    \frac{\R'_k}{\R_k} = \frac{\delta\phi'_k}{\delta\phi_k} - \frac{\pi'}{\phi'} \, .
\end{equation}
At super-Hubble scales, $\R_k$ freezes to a time-independent value: $\R'_k/\R_k \to 0$. In this limit, $\xi_\pi/\xi_\phi = \delta\phi'_k/\delta\phi_k = \pi'/\phi'$. The momentum and field kicks are proportional to the momentum and field time derivatives, respectively. In other words, the stochastic kicks induce only adiabatic fluctuations along the classical, non-stochastic background trajectory \cite{Tomberg:2022mkt}. This single-clock trajectory determines $\pi$ as a function of $\phi$, and the second equation in \eqref{eq:bg_eom} becomes redundant.

To describe motion along the classical trajectory, I define the function $\phieff$ as the field value on this trajectory at a given time $N$. Let $\Neff$ be the inverse of $\phieff$, that is, the time when a particular field value is reached, and let $\pieff=\phieff'$ be the derivative of $\phieff$. Below I'll abuse the notation so that, by default, $\phieff$ is evaluated at the current clock time $N$, but $\pieff$ is evaluated at $\Neff$, and $\Neff$ is evaluated at the current field value $\phi$, so that $\pieff$ is indeed a function of $\phi$, as alluded above. Switching $\pi$ to $\pieff$ in the first equation in \eqref{eq:bg_eom} corresponds to constraining the stochastic motion onto the classical trajectory. The equation becomes, term by term,
\begin{equation} \label{eq:phieff_eom}
    \dd \phi = \pieff \, \dd N + \sqrt{\Pphi(N,\kc) \dd N} \, \xiN \, ,
\end{equation}
where I switched to finite time steps of length $\dd N$ and renormed the noises to have unit variance, $\expval{\xiN\xiNd} = \delta_{NN'}$. I will switch between discrete and continuous time whenever it is beneficial for clarity.

Before solving equation~\eqref{eq:phieff_eom}, one must first solve $\pieff$ from the classical trajectory. I emphasize that even though we are left with a single first-order differential equation \eqref{eq:phieff_eom}, the full non-linear dynamics of \eqref{eq:bg_eom} are included through the pre-solved $\pieff$; the current procedure simply separates the problem into a classical part for $\pieff$ and a stochastic part for $\phi$. One must also solve $\Pphi(N,\kc)$ before using \eqref{eq:phieff_eom}. I will discuss solving both $\pieff$ and $\Pphi(N,\kc)$ in the next section. 

The stochastic process \eqref{eq:phieff_eom} is related to cosmological perturbations through the $\dN$ formalism \cite{Sasaki:1995aw, Sasaki:1998ug, Wands:2000dp, Lyth:2004gb}. We evolve $\phi$ starting from its classical value at $N_\text{ini}$ up to a fixed time $N$. The final field value corresponds to a classical e-fold number $\Neff$, now a stochastic variable. The initial and final times correspond to two coarse-graining scales, $\kini = \kc(N_\text{ini})$ and $k=\kc(N)$. The comoving curvature perturbation $\R$, coarse-grained at $k$, equals the difference between $N$ and $\Neff$ \cite{Tomberg:2022mkt}:
\begin{equation} \label{eq:Delta_N}
    \R_{<k} = \dN \equiv N -  \Neff \, .
\end{equation}
Indeed, through the noise in \eqref{eq:phieff_eom}, $\R_{<k}$ collects contributions from all perturbations with wavenumbers from $\kini$ to $k$, but not beyond. Repeating the stochastic evolution generates statistics for $\R_{<k}$.

\section{\texorpdfstring{$\bm{\dN}$}{Delta N} distribution with constant roll}
Equation \eqref{eq:phieff_eom} was solved numerically in \cite{Tomberg:2022mkt}. I will now solve it analytically for \emph{constant-roll inflation} \cite{Martin:2012pe, Motohashi:2014ppa}, defined by
\begin{equation} \label{eq:cr_eps}
    \epseff_2 \equiv \frac{\epseff'_1}{\epseff_1} = \text{const.} \, ,
    \quad \text{where} \quad
    \epseff_1 \equiv \frac{1}{2}\pieff^2 \ll 1 \, .
\end{equation}
Note that these slow-roll parameters are computed on the classical trajectory, as the tildes indicate. Below I'll drop the tilde over the constant $\eps_2$ to lighten the notation. I demand $\eps_2 > -3$ to ensure that super-Hubble perturbations freeze\footnote{This excludes, in particular, ultra-slow-roll inflation with $\eps_2 = -6$. As discussed in Section~\ref{sec:PBHs}, it is positive $\eps_2$ values that are important for PBH formation.} \cite{Tomberg:2022mkt}. From \eqref{eq:cr_eps}, we get
\begin{equation}
\label{eq:cr_field_rels}
\begin{gathered}
    \pieff' = \frac{\eps_2}{2}\pieff
    \quad \Rightarrow \quad
    \pieff = \frac{\eps_2}{2}\phieff \\
    \Rightarrow \,\,
    \phieff(N) = \phi_0 e^{\frac{\eps_2}{2}N} = \frac{2}{\eps_2}\sqrt{2\epseff_1(N)} \, , \quad
\end{gathered}
\end{equation}
where I integrated repeatedly. In the first step, I omitted the integration constant, corresponding to a shift in $\phi$; in the second step, I introduced the constant $\phi_0$, absorbed into the first slow-roll parameter. On the last line, I evaluated everything at $N$, but relations \eqref{eq:cr_field_rels} apply generically for the functions  $\phieff$, $\pieff$, and $\epseff_1$ at any parameter value.

As \eqref{eq:cr_field_rels} shows, $\pieff$ is linear in $\phi$ during constant roll\footnote{In our abused notation, $\pieff = \pieff(\Neff(\phi)) = \frac{\eps_2}{2}\phieff(\Neff(\phi)) = \frac{\eps_2}{2}\phi$.}. Also, as shown in \cite{Tomberg:2022mkt}, the stochasticity of the background does not affect the evolution of the short-wavelength perturbations during constant roll, and $\Pphi$ can be pre-computed in a non-stochastic background. It is then a known function of time, and we can write it as $\Pphi(N,\kc) = 2\epseff_1(N)\PR(\kc)$, where I dropped the time dependency of $\PR$, which we assume to be frozen. Then \eqref{eq:phieff_eom} is a first-order linear differential equation and can be solved via standard methods. The homogeneous equation with $\xi_N=0$ is solved by $\phi = \phieff(N)$, so the general solution is $\phi = \phieff(N) \times f(N)$. Substituting this into \eqref{eq:phieff_eom} and using \eqref{eq:cr_field_rels} yields
\begin{gather}
    \label{eq:phieff_solution}
    \phi(N) = \phieff(N)\qty[1 - \frac{\eps_2}{2}X(N)] \, , \\
    \label{eq:X_as_a_sum}
    X(N) \equiv -\!\sum_{k=\kini}^{\kc(N)} \sqrt{\PR(k) \, \dd \ln k} \, \xik \, .
\end{gather}
I delegated the solution's stochastic part to an auxiliary variable $X$, which follows the equation\footnote{Since the expectation value of $\dd X$ vanishes, this stochastic process is a \emph{martingale}. Many rigorous mathematical results exist for such processes, see e.g. \cite{martingales}.}
\begin{equation} \label{eq:X_eq}
    \dd X =
    - \sqrt{\PR(k)\,\dd \ln k} \, \xik \, ,
\end{equation}
and used the coarse-graining scale $\kc$ as a time variable indexing the steps, with $\dd N = \dd \ln k$ and $\expval{\xik\xikd} = \delta_{kk'}$. Note that all $\sigma$-dependency is lost; from now on, I drop the subscript $\sigma$ for brevity.

The integrated $X$ is a sum of independent Gaussian random variables, itself a Gaussian with mean zero and variance equal to the sum of the component variances. Its probability density is
\begin{equation} \label{eq:p_X}
    p[X(k)]=\frac{1}{\sqrt{2\pi}\sig{k}}e^{-\frac{X^2(k)}{2\sig{k}^2}} \, ,
    \quad
    \sig{k}^2 \equiv \int_{\kini}^k \PR(k') \, \dd \ln k' \, ,
\end{equation}
where I again passed to the continuum limit. Writing $\phi(N)=\phieff(\Neff)=\phi_0e^{\frac{\eps_2}{2}\Neff}$ and $\phieff(N)=\phi_0e^{\frac{\eps_2}{2}N}$ from \eqref{eq:cr_field_rels}, and using \eqref{eq:Delta_N}, we get a relationship between $X$ and $\dN$:
\begin{equation} \label{eq:X}
    X = \frac{2}{\eps_2}\qty(1-e^{-\frac{\eps_2}{2} \dN}) \, .
\end{equation}
The probability density of $\dN$ is then
\begin{equation} \label{eq:p_Delta_N}
\begin{aligned}
    p[\dN(k)] &= p[X(k)] \qty|\frac{\dd X}{\dd \dN}| \\
    &= \frac{1}{\sqrt{2\pi}\sig{k}} \exp\bigg[-\frac{2}{\sig{k}^2\eps_2^2}\qty(1-e^{-\frac{\eps_2}{2}\dN(k)})^2
    \\
    &\phantom{=\frac{1}{\sqrt{2\pi}\sig{k}} \exp X}- \frac{\eps_2}{2}\dN(k) \bigg] \, .
\end{aligned}
\end{equation}
I explicitly wrote out the $k$-dependence to emphasize that $\dN(k)$ gives $\R_{<k}$ coarse-grained over a specific scale $k$, adjustable by changing the upper limit of integration for $\sig{k}$ in \eqref{eq:p_X}.

\begin{figure}
    \centering
    \includegraphics{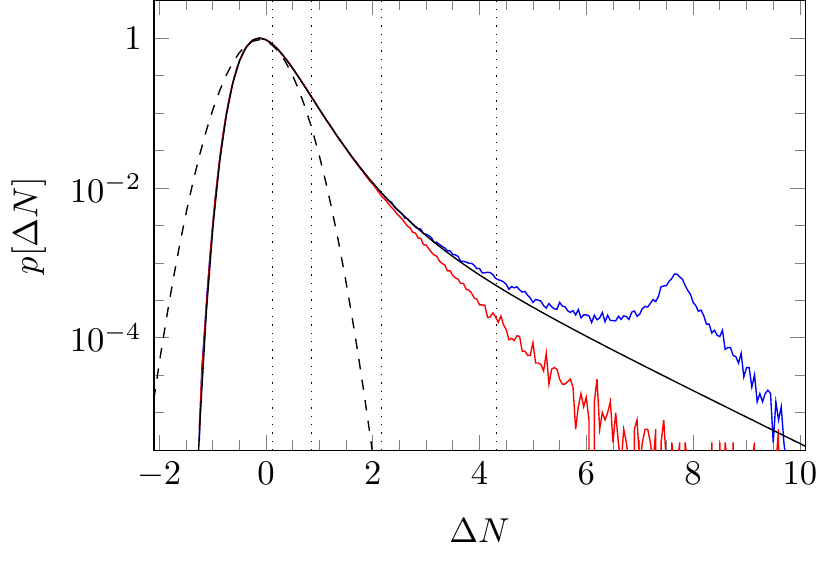}
    \caption{The probability distribution $p[\dN(k)]$: solved numerically from \eqref{eq:phieff_eom} ($10^7$ points in bins of width $0.05$) for an example model (upper blue line; see Appendix~\ref{sec:toy_model}) and its pure-constant-roll version (lower red line), the corresponding analytical result \eqref{eq:p_Delta_N} (middle black line), and a Gaussian approximation with variance $\sig{k}^2$ and mean matched to the distribution's peak (dashed line). Dotted lines, from left to right, correspond to $\eps_2 \Delta N(k) = 0.2$, $2\ln 2$, $2\ln \frac{2}{\sig{k}\eps_2}$, and $4\ln \frac{2}{\sig{k}\eps_2}$.}
    \label{fig:p}
\end{figure}

\begin{figure}
    \centering
    \includegraphics{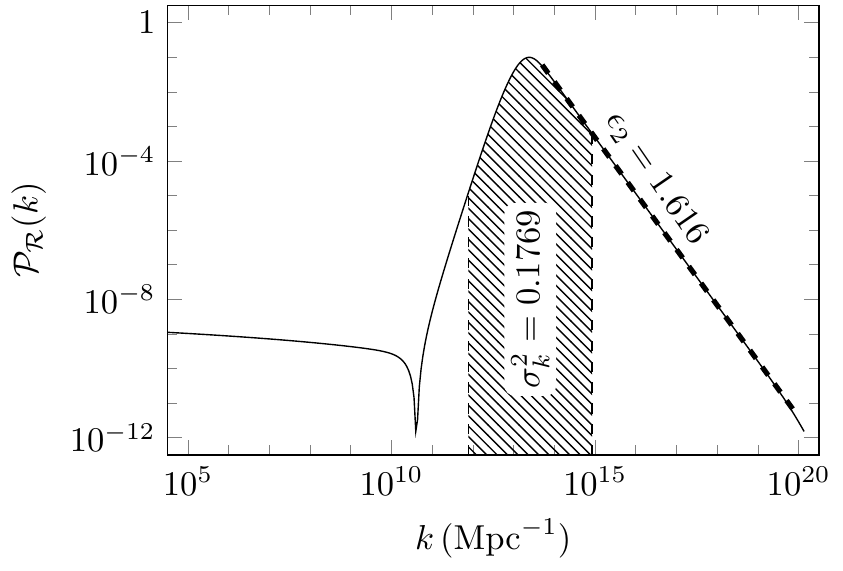}
    \caption{The power spectrum of the example model. The region between $\kini$ and $k_\text{PBH}$ is shaded. The parameter values are deduced from $\sig{k}^2=\int_{\kini}^{k_\text{PBH}} \PR(k) \, \dd \ln k$ and $\eps_2 = -\dd \ln \PR(k_\text{PBH}) / \dd \ln k$ (for the latter, see Section~\ref{sec:PBHs}).}
    \label{fig:PR}
\end{figure}

In Figure~\ref{fig:p}, I compare \eqref{eq:p_Delta_N} to a numerical solution of \eqref{eq:phieff_eom} for an example model from \cite{Tomberg:2022mkt} with a dominant constant-roll phase, see Appendix~\ref{sec:toy_model} for details. The power spectrum $\PR(k)$ is peaked at scales that exit the coarse-graining scale during the constant roll, see Figure~\ref{fig:PR}. I make a second comparison to a matching numerically solved pure constant-roll case, where \eqref{eq:cr_field_rels} applies exactly, but with the power spectrum from the above model. For small $\Delta N$, the match is excellent. In the $\dN \to 0$ limit, the distribution is Gaussian, with variance $\sig{k}^2$ and a slightly shifted mean. For large $\dN$, the distributions start to deviate. We see a non-Gaussian, exponential tail, familiar from previous studies of stochastic inflation \cite{Pattison:2017mbe, Ezquiaga:2019ftu, Prokopec:2019srf, Rigopoulos:2021nhv, Pattison:2021oen, Achucarro:2021pdh, Tada:2021zzj, Ahmadi:2022lsm, Animali:2022otk}.

To understand the large $\dN$ behaviour, note that $X$ has a boundary value $2/\eps_2$ at $\eps_2\dN \to \infty$, see \eqref{eq:X}, corresponding to $\phi=0$ in the convention of \eqref{eq:cr_field_rels}. This is an asymptotic limit of the classical trajectory: in constant-roll inflation, the field can only travel a finite distance. However, extreme $\xik$ fluctuations in \eqref{eq:X_eq} can take $X$ beyond this point\footnote{Similar behavior was studied in a simplified setup in \cite{Prokopec:2019srf}.}. For such fluctuations, results \eqref{eq:p_X} and \eqref{eq:p_Delta_N} become unreliable. I want these problems to take place in the tail of the probability distribution, at least one sigma away from the mean; to this end, I require $\sig{k} \lesssim |2/\eps_2|$, or $\sig{k}|\eps_2| \lesssim 2$. In addition, for the results to be valid at a given $X(k)$, I demand the distance to $2/\eps_2$ to be at least $\sig{k}$, that is, $\eps_2X(k) \lesssim 2 - \sig{k}|\eps_2|$, or
\begin{equation} \label{eq:N_lim}
    \eps_2 \dN \lesssim 2\ln\frac{2}{\sig{k}|\eps_2|} \, .
\end{equation}
Beyond this, the stochastic motion of our example model starts to probe dynamics beyond the constant-roll phase, and the results become unreliable since the pre-computed $\Pphi$ can no longer be trusted. In the pure constant-roll case, trajectories beyond \eqref{eq:N_lim} diverge. In the numerics, this happened to three points in a thousand; these were discarded from Figure~\ref{fig:p}, making the pure constant-roll result also unreliable beyond \eqref{eq:N_lim}. Note also that because of the $\eps_2 X < 2$ limitation, the distributions \eqref{eq:p_X}, \eqref{eq:p_Delta_N} are not properly normalized; however, in our limit of small $\sig{k}|\eps_2|$, the correction is negligible.

The transition to the exponential tail starts approximately when the exponent shifts from concave to convex behavior, that is $\partial^2_{\Delta N}\ln p = 0$, and completes when $\-\frac{\eps_2}{2}$ dominates the exponent's derivative $\partial_{\Delta N} \ln p$. Solving these conditions from \eqref{eq:p_Delta_N} we get the following hierarchy:
\begin{itemize}
    \item $|\eps_2\Delta N(k)| \gtrsim 0.1 \times 2$: Gaussianity of $\Delta N$ fails,
    \item $\eps_2 \Delta N(k) \approx 2\ln 2$: Transition to exponential tail starts,
    \item $\eps_2 \Delta N(k) \gtrsim 2\ln \frac{2}{\sig{k}|\eps_2|}$: $X$ too close to $2/\eps_2$; results \eqref{eq:p_X}, \eqref{eq:p_Delta_N} fail,
    \item $\eps_2\Delta N(k) \gtrsim 4\ln \frac{2}{\sig{k}|\eps_2|}$: Transition to exponential tail completes.
\end{itemize}
The exponential tail never comes to fully dominate since the limit \eqref{eq:N_lim} is reached first and our single-clock stochastic description fails.

For completeness, I compute the expectation value of $\dN(k)$ by inverting \eqref{eq:X}:
\begin{equation} \label{eq:dN_bar}
\begin{aligned}
    \bdN(k) &\approx  -\int_{-\infty}^{\infty} \frac{2}{\eps_2}\ln \hspace{-2pt} \qty[1-\frac{\eps_2}{2}X(k)]  p[X(k)] \dd X(k) \\
    &= \sum_{n=1}^{\infty} \frac{(2n-1)!}{n!}\frac{\sig{k}^{2n}\eps_2^{2n-1}}{2^{3n - 1}} \, ,
\end{aligned}
\end{equation}
where I Taylor expanded the logarithm around $X(k)=0$. The sum diverges as the factorial in the numerator grows; this is due to the improper bounds I used in the integral, going beyond $X=2/\eps_2$. However, for small $\sig{k}|\eps_2|$, keeping only the first terms in the sum provides an excellent approximation for the properly bounded integral. The result is negligibly small. This is good: in \eqref{eq:Delta_N}, $\dN$ was defined as the difference between the elapsed time and the non-stochastic background time, while technically, $\R_{<k}$ is better given by the difference between the elapsed time and its mean, equivalent to $\dN - \bdN$. The difference is not significant and will be neglected in this paper. Note also that $\eps_2\bdN(k)$ is positive due to the heavy tail at $\eps_2\dN>0$, but the peak of the distribution \eqref{eq:p_Delta_N} actually has $\eps_2\dN_\text{peak}<0$, given by
\begin{equation}
    \dN_\text{peak} = -\frac{2}{\eps_2}\ln\qty[\frac{1}{2}\qty(1+\sqrt{1+\sig{k}^2\eps_2^2})] \, .
\end{equation}

\section{Stochastic trajectories}
Let us study more closely the stochastic trajectories that end at a fixed $X(k)$, and thus a fixed $\dN(k)$. At an intermediary scale $k'$, $\kini < k' < k$, the conditional probability density of $X$ is
\begin{equation} \label{eq:p_X_cond}
    p[X(k') | X(k)] = \frac{p[X(k')]\,p[X(k')\to X(k)]}{p[X(k)]} \, ,
\end{equation}
where the transition probability from $X(k')$ to $X(k)$ is
\begin{equation} \label{eq:p_X_trans}
\begin{aligned}
    p[X(k')\to X(k)] =
    \frac{\exp{-\frac{\qty[X(k')-X(k)]^2}{2(\sig{k}^2 - \sig{k'}^2)}} }{\sqrt{2\pi\qty(\sig{k}^2 - \sig{k'}^2)}}\, .
\end{aligned}
\end{equation}
The division by $p[X(k)]$ ensures the correct normalization, $\int_{-\infty}^{\infty} p[X(k') | X(k)] \dd X(k') = 1$. Using \eqref{eq:p_X} and \eqref{eq:p_X_trans}, \eqref{eq:p_X_cond} becomes
\begin{equation} \label{eq:p_X_cond_2}
\begin{gathered}
    p[X(k') | X(k)] = \frac{1}{\sqrt{2\pi}\sigC{k'}{k}}e^{-\frac{\qty[X(k')-\overbar{X}(k')_{|k}]^2}{2\sigC{k'}{k}^2}} \, , \\
    \sigC{k'}{k}^2 \equiv \frac{\sig{k'}^2}{\sig{k}^2}\qty(\sig{k}^2 - \sig{k'}^2) \, , \qquad
    \overbar{X}(k')_{|k} \equiv \frac{\sig{k'}^2}{\sig{k}^2} X(k) \, .
\end{gathered}
\end{equation}
Stochastic trajectories ending at $X(k)$ cluster around the `most probable path' given by the conditional expectation values $\overbar{X}(k')_{|k}$.

The corresponding expectation values for $\dN(k')$ are
\begin{equation} \label{eq:dN_bar_cond}
\begin{aligned}
    \bdN(k')_{|k} &\equiv  -\int_{-\infty}^{\infty} \frac{2}{\eps_2}\ln \hspace{-2pt} \qty[1-\frac{\eps_2}{2}X(k')] \\
    &\qquad\qquad \times p[X(k') | X(k)] \, \dd X(k') \\
    &= -\frac{2}{\eps_2}\ln \hspace{-2pt}\qty[1-\frac{\eps_2}{2}\overbar{X}(k')_{|k}] \\
    &\phantom{=} + \sum_{n=1}^{\infty} \frac{(2n-1)!}{2^{n-1}n!}\frac{\sigC{k'}{k}^{2n}\eps_2^{2n-1}}{\qty[2-\eps_2\overbar{X}(k')_{|k}]^{2n}} \, ,
\end{aligned}
\end{equation}
analogously to \eqref{eq:dN_bar}, where this time I expanded around $X(k')=\overbar{X}(k')_{|k}$. Again, the first terms in the formally divergent sum give a good approximation.

In \cite{Tomberg:2022mkt}, $\bdN(k')_{|k}$ was approximated by minimizing an exponent in the probability density of the stochastic noise. In constant roll, \eqref{eq:dN_bar_cond} is a strict improvement over the method of \cite{Tomberg:2022mkt} since it also takes into account the integration volume and reproduces the expectation value accurately. Indeed, in Appendix~\ref{sec:probable_paths} I show that \cite{Tomberg:2022mkt} only reproduces the $\sigC{k'}{k} \to 0$ limit of \eqref{eq:dN_bar_cond}. In \cite{Tomberg:2022mkt}, the most probable paths in $\Delta N$ were used as an ideal bias for importance sampling, see also \cite{Jackson:2022unc}, allowing efficient numerical studies of the far tail of the probability distribution. For numerical studies in constant roll, I recommend working in $X$ and converting to $\Delta N$ only in the end to access $\R_{<k}$. With \eqref{eq:p_X_cond_2}, the ideal bias for the noise term $\xikd$ is
\begin{equation} \label{eq:ideal_bias}
    \xioverbar{\xi}_{k'} = -\frac{\dd \overbar{X}(k')_{|k}}{\sqrt{\PR(k')\,\dd \ln k}} = -\sqrt{\PR(k')\,\dd \ln k} \, \frac{X(k)}{\sig{k}^2} \, ,
\end{equation}
which is the largest for the modes with the strongest $\PR(k')$.

With the typical trajectories \eqref{eq:p_X_cond_2}--\eqref{eq:dN_bar_cond} at hand, we may take a second look at the region of applicability of the original probability distributions \eqref{eq:p_X}, \eqref{eq:p_Delta_N}. The distributions are reliable for a given $X(k)$ as long as the typical paths don't go too close to the boundary value $X = 2/\eps_2$. From \eqref{eq:p_X_cond_2}, the variances on the paths are of order $\sig{k}$, so good paths have $\eps_2X(k) \lesssim 2 - \sig{k}|\eps_2|$. This excludes the same $\dN$ region as above in \eqref{eq:N_lim}; applying the requirement at the distribution mean $X\approx0$ reproduces our old limit $\sig{k}|\eps_2| \lesssim 2$. Knowledge of the most probable paths has allowed us to better understand these limits. 

We may also ask what dominates the typical evolution: the classical drift or the stochastic kicks. These correspond to the first and second terms on the right-hand side of equation~\eqref{eq:phieff_eom}, respectively, with the typical noise given by \eqref{eq:ideal_bias}. For the classical drift to dominate, the ratio of the terms should be larger than one in magnitude. After some algebra involving \eqref{eq:cr_field_rels}, \eqref{eq:X}, and \eqref{eq:p_X_cond_2}, this condition can be expressed as $\big|\frac{\dd}{\dd \ln k'} \bdN(k')_{|k}\big| < 1$, where $\bdN(k')_{|k}$ is the leading order expression from \eqref{eq:dN_bar_cond}; in other words, the evolution is classical if $\dN$ grows slower than $N$ (remember that $\dd \ln k = \dd N$). Expanding \eqref{eq:X} in different $\dN(k)$ regimes, we can also estimate the condition as:
\begin{itemize}
    \item $|\eps_2\dN(k)| \ll 1$: Trajectories are classical for $\PR(k') < \sig{k}^2/|\dN(k)|$. 
    \item $\eps_2\dN(k) \ll -1$: Trajectories are classical for $\PR(k') < |\eps_2|\sig{k'}^2/2$.
    \item $\eps_2\dN(k) \gg 1$: The stochastic evolution takes $X$ beyond the classical trajectory and the current analysis breaks down. Trajectories are never classical.
\end{itemize}
These improve the approximate condition $\PR(k') \ll 1$ commonly used in the literature, see e.g. \cite{Vennin:2015hra} for a discussion.

\section{Implications for primordial black hole production}
\label{sec:PBHs}
As discussed above, a typical single-field, PBH-producing model of inflation proceeds through two phases: first an ultra-slow-roll phase with $\eps_2 \lesssim -6$, and then a dual constant-roll phase with $\eps_2 \gtrsim 0$ \cite{Karam:2022nym}. The power spectrum $\PR(k)$ is enhanced for scales that exit the Hubble radius during ultra slow roll. However, the coarse-graining parameter $\sigma$ introduces a delay between the Hubble and coarse-graining exits of a given mode. The most enhanced modes get coarse-grained only during the constant-roll phase, so that's when the stochastic effects are strongest and the PBH-producing non-linear perturbations develop \cite{Tomberg:2022mkt}. Thus, the constant-roll formalism derived above is the right tool to compute PBH statistics.

For simplicity, I assume that all patches exceeding a collapse threshold of $\R_c = \dN_c \approx 1$ collapse into PBHs \cite{Green:2004wb, Harada:2013epa, Escriva:2019phb} when the corresponding scale $k$ re-enters the Hubble radius after inflation. The PBH is formed from all mass within the Hubble radius, with approximate mass\footnote{I assume standard expansion history, where PBHs form early in the radiation-dominated era, and the Hubble parameter today is $H_0 \approx 70 \text{km/s/Mpc}$.}
\begin{equation} \label{eq:PBH_mass}
    M_k \approx 6.8 \times 10^{49}\mathrm{g} \, \qty(\frac{k}{0.05 \, \text{Mpc}^{-1}})^{-2} \, .
\end{equation}
The PBH mass fraction today is
\begin{equation} \label{eq:abundance_master}
    \Omega_\text{PBH} \approx 6.2\times10^{25} \beta \,\qty(\frac{M_k}{\mathrm{g}})^{-1/2} \, ,
\end{equation}
where $\beta$ is their mass fraction when they form. The enhancement is a result of PBH density diluting slower than the radiation energy density during radiation domination. To compute $\beta$, the probability density $p(\dN)$ is typically assumed to be Gaussian, leading to\footnote{The prefactor of two is a common correction employed in the Press--Schechter formalism \cite{Green:2004wb}.} \cite{Green:2004wb}
\begin{equation} \label{eq:abundance_Gaussian}
    \beta_k = 2\int_{\dN_c}^{\infty} \frac{\dd \dN}{\sqrt{2\pi}\sig{k}}e^{-\frac{\dN^2}{2\sig{k}^2}} = \erfc\qty[\frac{\dN_c}{\sqrt{2}\sig{k}}] \, .
\end{equation}
With our more careful analysis, \eqref{eq:p_X}--\eqref{eq:p_Delta_N} give
\begin{equation} \label{eq:beta}
\begin{aligned}
    \beta_k &= 2\int_{X_c}^{2/\eps_2} \dd X(k) \, p[X(k)] \\
    &= \erfc\qty[\frac{X_c}{\sqrt{2}\sig{k}}] - \erfc\qty[\frac{2/\eps_2}{\sqrt{2}\sig{k}}]  \\
    &\approx \frac{\sig{k}\eps_2}{\sqrt{2\pi}}\frac{\exp[-\frac{2}{\sig{k}^2\eps_2^2}\qty(1-e^{-\frac{\eps_2}{2}\dN_c})^2]}{1-e^{-\frac{\eps_2}{2}\dN_c}} \, ,
\end{aligned}
\end{equation}
where $X_c$ is related to $\dN_c$ by \eqref{eq:X}, I approximated the second complementary error function as $0$, corresponding to sending the upper integration limit to infinity, and I used a large-argument approximation for the first error function. The result merges with the Gaussian approximation \eqref{eq:abundance_Gaussian} in the $\eps_2 \to 0$ limit. As discussed around \eqref{eq:N_lim}, this result is reliable for $\eps_2\sig{k}\lesssim 2$, $\Delta N_c \lesssim \frac{2}{\eps_2}\ln \frac{2}{\sig{k}\eps_2}$. These are comfortably satisfied for $\dN_c \approx 1$, small $\sig{k}$, and $\eps_2$ at most of order one. PBH production typically takes place in the transition region between the Gaussian middle and the large-$\dN$ exponential tail of the probability distribution.

Let us next find the parameter space that produces a desired PBH abundance $\beta$. From \eqref{eq:X} and \eqref{eq:beta}, we have
\begin{equation} \label{eq:eps2_sigma_relation}
    \frac{\sqrt{2}}{\sig{k}\eps_2}\qty(1-e^{-\frac{\eps_2}{2}\Delta N_c}) \approx \erfc^{-1}(\beta_k) \equiv E \, ,
\end{equation}
linking $\sig{k}$ and $\eps_2$ together. Equation~\eqref{eq:N_lim} limits their variation to
\begin{equation} \label{eq:eps2_sigma_limist}
\begin{aligned}
    \eps_2 &= 0 \ \dots \ \frac{2}{\Delta N_c}\ln(1+\sqrt{2}E) \, , \\
    \sig{k} &= \frac{\Delta N_c}{(\sqrt{2}E +1)\ln(\sqrt{2}E +1)} \ \dots \ \frac{\Delta N_c}{\sqrt{2}E}  \, .
\end{aligned}
\end{equation}
The $\eps_2=0$ and high $\sig{k}$ values correspond to slow-roll inflation with Gaussian statistics; the high $\eps_2$ and low $\sig{k}$ values correspond to a maximal stochastic effect still reliably captured by our method.
For example, if PBHs of mass $10^{19} \, \mathrm{g}$---in the asteroid mass window \cite{Green:2020jor, Carr:2021bzv}---constitute all dark matter, $\Omega_\text{PBH} \approx 0.27$, then by \eqref{eq:abundance_master} $\beta\approx 1.4\times 10^{-17}$ and $E \approx 6.0$. With $\Delta N_c \approx 1$, \eqref{eq:eps2_sigma_limist} gives $\eps_2 = 0 \dots 4.5$, $\sig{k} = 0.05 \dots 0.12$. Introducing a non-zero $\eps_2$ allows us to decrease $\sig{k}$ by $60\%$, corresponding to a drop of $84\%$ in $\PR$, say, from $\PR\sim 0.01$ to $0.0016$. This way, stochastic effects slightly alleviate the fine-tuning needed to produce PBHs from a high $\PR$ peak.

To provide another point of view, fixing $\sig{k}=0.05$ in the above scenario, the slow-roll parameter is allowed to vary as $\eps_2 = 0\dots 3.3$, giving (naively) $\Omega_\text{PBH} = 0.27\dots 10^{11}$. A model that is tuned correctly in the Gaussian approximation can severely overproduce PBHs when stochastic effects are taken into account.

In \eqref{eq:beta}, the PBH abundance is given in terms of two parameters, $\sig{k}$ and $\eps_2$. The first of these is an integral over the power spectrum peak, see \eqref{eq:p_X}; however, the second one can also be deduced from the power spectrum. As shown in \cite{Karam:2022nym}, in the standard setup of ultra slow roll followed by constant roll, the power spectrum decays like $\PR(k) \propto k^{-\eps_2}$ after the peak. Computing the stochastic abundance \eqref{eq:beta} is then no more demanding than computing the Gaussian estimate: for both, knowledge of the power spectrum is enough. Figure~\ref{fig:PR} demonstrates this for the example model. I have here focused on PBHs of a fixed mass; to get the PBH statistics in different mass bins, it is enough to vary the $k$ in $\sig{k}$.

In Appendix~\ref{sec:higgs_models}, I compare results \eqref{eq:p_Delta_N} and \eqref{eq:beta}, starting from the power spectrum, to heavy-duty numerical computations of the full stochastic system \eqref{eq:bg_eom} of modified Higgs potentials from \cite{Figueroa:2021zah}. The match is excellent.

\section{Comparison to non-stochastic \texorpdfstring{$\bm{\dN}$}{Delta N} computations}
PBH formation with constant roll was considered earlier in \cite{Cai:2018dkf, Atal:2019cdz, Atal:2019erb, Biagetti:2021eep, Hooshangi:2021ubn, Hooshangi:2022lao, Pi:2022ysn, Firouzjahi:2023ahg}, with the $\dN$ formalism but without stochasticity. These studies start with an initial Gaussian field perturbation $\Delta \phi$ and let the non-linear, classical background dynamics turn it into a final $\Delta N$. Using this strategy, \cite{Pi:2022ysn} derived a probability distribution\footnote{Perturbations active during ultra slow roll, $\eps_2 = -6$, follow a dual distribution with $p \sim e^{-3\Delta N}$ \cite{Biagetti:2021eep, Pi:2022ysn}.} identical to \eqref{eq:p_Delta_N}. In this paper's notation,
\begin{equation} \label{eq:Delta_phi}
\begin{split}
    \Dphi(N) &\equiv \phi(N) - \phieff(N) = -\pieff(N)X(N) \\
    &= \sqrt{2\epseff_1(N)}\frac{2}{\eps_2}\qty(1-e^{-\frac{\eps_2}{2}\dN})\, ,
\end{split}
\end{equation}
where I used \eqref{eq:cr_field_rels}, \eqref{eq:phieff_solution}, and \eqref{eq:X}. Since the stochastic equation \eqref{eq:phieff_eom} is linear in $\phi$ during constant roll, $\Dphi$ separates into the stochastic $X$ and a non-stochastic prefactor, and is, indeed, Gaussian, inheriting this feature from $X$. It does not matter when the stochastic kicks in $X$ are applied: an initial, macroscopic kick of sufficient strength produces the same result as microscopic kicks distributed over time. The former, the method of \cite{Cai:2018dkf, Atal:2019cdz, Atal:2019erb, Biagetti:2021eep, Hooshangi:2021ubn, Hooshangi:2022lao, Pi:2022ysn, Firouzjahi:2023ahg}, successfully reproduces the exponential $\dN$ tail in constant roll.

I have provided a stochastic interpretation for this method and clarified the time and scale dependence of the process. In particular, the later constant-roll phase turns out to be dynamically more important than the earlier ultra-slow-roll phase. It was already pointed out in \cite{Biagetti:2021eep, Hooshangi:2021ubn} that PBHs form from the transition region between the Gaussian part and the exponential tail of the probability distribution; I confirmed this and pointed out that the tail is, in fact, not reliably resolved by current methods.\footnote{I thank Andrew Gow for discussions on the properties of the far tail of the distribution and its relevance for PBH formation.}
Finally, the approach of \cite{Cai:2018dkf, Atal:2019cdz, Atal:2019erb, Biagetti:2021eep, Hooshangi:2021ubn, Hooshangi:2022lao, Pi:2022ysn, Firouzjahi:2023ahg} breaks down outside of constant-roll (potentially even if multiple regions of constant roll are combined, such as in \cite{Pi:2022ysn}), where the stochastic equation \eqref{eq:phieff_eom} is no longer linear. Outside of constant roll, there is no reason to assume $\Dphi$ to be fully Gaussian, or the method of a strong initial kick to reproduce a more accurate stochastic result.

\section{Conclusions}
In this paper, I used the stochastic $\dN$ formalism to compute the probability distribution of the coarse-grained curvature perturbation $\R$ in constant-roll inflation, starting from the linear power spectrum $\PR$. I then studied the formation probability of primordial black holes. PBH production is controlled by a transition region between the probability distribution's Gaussian center and exponential tail.

The results apply very generally to all PBH-producing models of single-field inflation where the field rolls over a local maximum of its potential, or, equivalently, where an ultra-slow-roll phase is followed by a dual constant-roll phase. The perturbation probability distribution is an analytical function of quantities derived from the power spectrum only, and thus using the improved, stochastic result \eqref{eq:beta} is not computationally any more expensive than the Gaussian approximation \eqref{eq:abundance_Gaussian}. The power spectrum can be pre-computed (beyond the de Sitter approximation) and used as an input for the stochastic evolution since there is no backreaction between the short-wavelength perturbations and the stochastic background during constant roll.

Recently, a series of papers \cite{Kristiano:2022maq, Kristiano:2023scm, Firouzjahi:2023aum, Motohashi:2023syh} have suggested that a high power spectrum $\PR\sim 10^{-2}$ may produce strong loop corrections that ruin the model's cosmic microwave background (CMB) predictions, ruling out inflationary PBHs (see \cite{Riotto:2023hoz, Riotto:2023gpm, Firouzjahi:2023ahg, Franciolini:2023lgy} for a critical take). I have demonstrated that stochastic effects can lower the $\PR$ required for PBH production, down to $\sim10^{-3}$ in an example case of constant-roll inflation. This helps alleviate the reported tension and demonstrates that one needs to go beyond traditional perturbation theory to make definitive claims about the compatibility of the CMB and inflationary PBHs.

The PBH mass and abundance formulas \eqref{eq:PBH_mass}, \eqref{eq:abundance_master} are approximations commonly used in the literature. In truth, the masses follow a distribution around \eqref{eq:PBH_mass}, and the abundance is better computed from the \emph{compaction function} \cite{Shibata:1999zs, Harada:2015yda, Musco:2018rwt, Tada:2021zzj, Ferrante:2022mui, Gow:2022jfb}, a quantity related but not identical to $\R_k$. I leave a compaction-function-based treatment for future work.

\begin{acknowledgments}
I thank Daniel Figueroa, Sami Raatikainen, Syksy R\"{a}s\"{a}nen, Joe Jackson, and Andrew Gow for discussions. I acknowledge the support of the European Consortium for Astroparticle Theory in the form of an Exchange Travel Grant. This work was supported by the Estonian Research Council grant PRG1055 and by the EU through the European Regional Development Fund CoE program TK133 ``The
Dark Side of the Universe.''
\end{acknowledgments}

\appendix

\section{Example model}
\label{sec:toy_model}
The example model of Figures~\ref{fig:p} and~\ref{fig:PR} is lifted from \cite{Tomberg:2022mkt} (the `Hubble-tailored model'). It is built starting from an analytical form of the first slow-roll parameter as a function of the number of e-folds $N$,
\begin{equation} \label{eq:Hubble_tailored_eps}
\begin{aligned}
    \eps_1 =& \, \eps_{1,\text{top}}
    \times
    e^{-3(N-N_1)}\frac{\cosh^2[\lambda(N_2-N)]}{\cosh^2[\lambda(N_2-N_1)]} \\
    &\times
    \qty[\frac{2}{1+e^{-\theta_\text{cut}(\lambda + 3/2)(N-N_1)}}]^{2/\theta_\text{cut}} \\
    &\times \qty[\frac{1+\sqrt{\alpha}\qty(\theta_\text{SR}^{-1}\ln 2)^{\beta/2}}{1+\sqrt{\alpha}\qty(\theta_\text{SR}^{-1}\ln\hspace{-0.1cm}\qty[1+e^{-\theta_\text{SR}(N-N_1)}])^{\beta/2}}]^2 \, .
\end{aligned}
\end{equation}
The different parts set $\eps_1$ to describe slow roll at early times, to shift to ultra slow roll with $\eps_2 \approx -3-2\lambda$ around $N_1$, and to shift to the dual constant roll with $\eps_2 \approx -3+2\lambda$ around $N_2$. Behavior \eqref{eq:Hubble_tailored_eps} determines everything about the model, including the underlying inflationary potential and the ensuing perturbation power spectrum $\PR(k)$, up to an overall energy scale that is fixed by CMB observations. I solve the power spectrum numerically from the Sasaki--Mukhanov equation, beyond the de Sitter approximation $\PR\approx \frac{H^2}{8\pi^2\eps_1}$. For more details, see Appendix~A in \cite{Tomberg:2022mkt}.

I use the same parameter values as \cite{Tomberg:2022mkt},
\begin{equation} \label{eq:eff_model_parameter_values}
\begin{gathered}
    \eps_{1,\text{top}} = 0.01 \, , \quad
    N_1 = 32 \, , \quad
    N_2 = 35.04 \, , \quad
    \lambda = 2.308 \, ,
    \\
    \alpha = 50 \, , \quad
    \beta = 1.28 \, , \quad
    \theta_\text{cut} = 1 \, , \quad
    \theta_\text{SR} = 5 \, ,
\end{gathered}
\end{equation}
where the e-folds are counted from the CMB scale and inflation ends ($\eps_1=1$) at $N=50$. The model is compatible with the latest CMB bounds \cite{Planck:2018jri, BICEP:2021xfz} and produces a power spectrum peak giving PBHs around the asteroid mass window \cite{Green:2020jor, Carr:2021bzv}. For easier numerics, the model is tuned to have strong stochastic effects---the power spectrum peak is $\PR(k_\text{peak})\sim 0.1$, and I evaluate $p[\dN(k_\text{PBH})]$ for $k_\text{PBH} \gg k_\text{peak}$ to include stochastic kicks over a wide range of scales, see Figure~\ref{fig:PR}. The relevant parameter values are
\begin{equation} \label{eq:eff_model_PR_parameters}
\begin{gathered}
    \kini = 7.725\times10^{11} \, , \quad
    k_\text{PBH} = 8.472\times10^{14}  \, , \\
    \sig{k=k_\text{PBH}}^2 \approx 0.1769 \, , \quad
    \eps_2 \approx 1.616 \, .
\end{gathered}
\end{equation}
The ensuing probability distribution in Figure~\ref{fig:p} overproduces PBHs but clearly shows the Gaussian, exponential, and transition regions described in the text.

\section{Most probable paths}
\label{sec:probable_paths}
In \cite{Tomberg:2022mkt}, equation \eqref{eq:phieff_eom} was rewritten in terms of the classical e-fold number $\Neff$, a proxy for $\phi$, as
\begin{equation} \label{eq:neff_eom_old}
    \dd \Neff = \dd N + \sqrt{\frac{\Pphi(N,\kc)}{2\epseff_1(\Neff)} \dd N} \, \xiN \, 
\end{equation}
(up to the slightly different $\kc$ convention, see footnote~\ref{ft:k} above). Working in $\Neff$ is appealing since it makes accessing $\dN = N - \Neff$ easy. However, \eqref{eq:neff_eom_old} is slightly inaccurate: it neglects a term arising from the non-linear relation between $\phieff$ and $\Neff$ and the stochastic nature of the differential equation. The correct form, given by \Ito's lemma \cite{Ito:1944, Vennin:2015hra}, can be arranged into
\begin{equation} \label{eq:neff_eom}
    {\dd \dN} = \frac{\Pphi(N,\kc)}{4\epseff_1(\Neff)}\frac{\epseff_2(\Neff)}{2} \dd N - \sqrt{\frac{\Pphi(N,\kc)}{2\epseff_1(\Neff)} \dd N} \, \xiN \, .
\end{equation}
The smallness of the power spectrum suppresses the first, extra term, at least for small perturbations. When computing the probability distribution up to $\dN \sim 1$, the difference between \eqref{eq:neff_eom_old} and \eqref{eq:neff_eom} is negligible for the modified Higgs model of \cite{Tomberg:2022mkt}, but noticeable for the Hubble-tailored one.

\begin{figure*}
    \centering
    \includegraphics{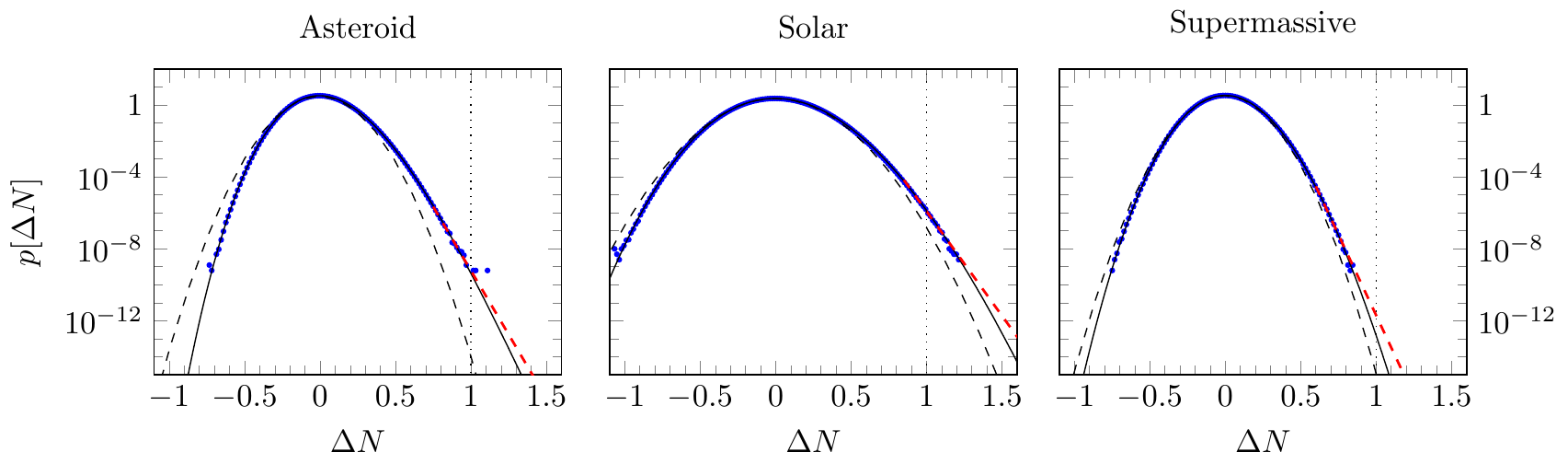}
    \caption{Probability distributions $p[\dN(k)]$ in the three models considered in \cite{Figueroa:2020jkf, Figueroa:2021zah} for asteroid mass, solar mass, and supermassive black holes: the numerical results from \cite{Figueroa:2021zah} (blue dots), exponential fits to their tails (red dashed lines), analytical results \eqref{eq:p_Delta_N} (solid black lines), and Gaussian approximations (dashed black lines). The dotted lines mark the collapse threshold $\dN_c=1$.}
    \label{fig:p3}
\end{figure*}

\begin{table*}
\begin{center}
\begin{tabular}{L{2.5cm} C{2cm} C{2.8cm} C{1.5cm} C{1.5cm} C{2cm} C{2cm} C{1cm}}
\toprule
& $M$ & $k_\text{PBH}$ & $\sig{k}$ & $\eps_2$ & $\beta_\text{num}$ & $\beta_\text{CR}$ & $\dN_\text{lim}$ \\
\midrule
Asteroid & $1.4\times10^{19} \, \mathrm{g}$ & $2.0\times10^{13}\,\mathrm{Mpc}^{-1}$ & $0.1232$ & $0.799$ & $3.7\times10^{-11}$ & $2.2\times10^{-11}$ & $7.5$ \\
Solar & $4.7\, M_\odot$ & $8.0\times10^5\,\mathrm{Mpc}^{-1}$ & $0.1748$ & $0.284$ & $9.0\times10^{-8}$ & $9.6\times10^{-8}$ & $26.0$ \\
Supermassive & $1.8\times10^3\,M_\odot$ & $3.9\times10^4\,\mathrm{Mpc}^{-1}$ & $0.11877$ & $0.292$ & $9.7\times10^{-14}$ & $0.5\times10^{-14}$& $27.7$ \\
\bottomrule
\end{tabular}
\end{center}
\caption{Various parameters for the three models considered in \cite{Figueroa:2020jkf, Figueroa:2021zah}, see Table~1 of \cite{Figueroa:2021zah}. Note that the formula \cite{Figueroa:2021zah} used to relate the black hole mass $M$ to the wavenumber $k$ differs slightly from \eqref{eq:PBH_mass}. $M_\odot\approx 2.0\times 10^{33} \, \mathrm{g}$ is the solar mass. The variables $\eps_2$ and $\sig{k}$ were computed from the power spectrum, giving $\beta_\text{CR}$ by the constant roll result \eqref{eq:beta}. The collapse fraction $\beta_\text{num}$ was computed from the exponential fit to the tail of the numerical distribution. $\dN_\text{lim}$ is the validity limit of the constant-roll approach from \eqref{eq:N_lim}, well over the collapse threshold in all cases.}
\label{tab:higgs_models}
\end{table*}

The $\Neff$ equation can be used to estimate the most probable path for a fixed final $\Delta N$. Following \cite{Tomberg:2022mkt}, we start from the probability distribution of the noises $\xiN$:
\begin{equation} \label{eq:xi_distr}
    p(\hat{\xi}) = \frac{1}{\sqrt{2\pi}}e^{-\frac{1}{2}\sum_{N}\xiN^2} \, .
\end{equation}
Using \eqref{eq:neff_eom} and going to the continuum limit, the exponent becomes
\begin{gather} \label{eq:xi_sum}
    -\frac{1}{2}\sum_{N}\xiN^2 = -\int \frac{(\Neff' - 1 + D)^2}{\Pphi(N)/\epseff_1(\Neff)}\dd N \, , \\
    D\equiv \frac{\Pphi(N)\epseff_2(\Neff)}{8\epseff_1(\Neff)} \, ,
\end{gather}
where the $\kc(N)$ argument of $\Pphi$ is implied and $D$ is the additional contribution from \Ito's lemma.
Minimizing this with respect to $\Neff(N)$ gives the differential equation
\begin{equation} \label{eq:Neff_path_eq}
\begin{split}
    \Neff'' - \frac{\epseff_2(\Neff)}{2}\qty(1-\Neff'^2 - D^2) + \frac{\Pphi'(N)}{\Pphi(N)}\qty(1-\Neff') \\
    + \, \epseff_3(\Neff)D(1-D) = 0 \, ,
\end{split}
\end{equation}
where $\epseff_3(\Neff)\equiv\epseff'_2(\Neff)/\epseff_2(\Neff)$. This is equation~(3.22) from \cite{Tomberg:2022mkt}, up to the different $\kc$ convention and the extra $D$ terms. For a direct comparison to \cite{Tomberg:2022mkt}, I will now approximate $D\approx0$. Going from $\Neff$ to $\dN$ with \eqref{eq:Delta_N} and rearranging, we then get
\begin{equation} \label{eq:dN_path_eq}
    \frac{\dN''}{\dN'} = \frac{\epseff'(\Neff)}{2\epseff(\Neff)}\dN' + \frac{\Pphi'(N)}{\Pphi(N)} - \frac{\epseff'(\Neff)}{\epseff(\Neff)} \, .
\end{equation}      
In constant roll, $\epseff'(\Neff)/\epseff(\Neff)=\eps_2$ is a constant, and we can integrate \eqref{eq:dN_path_eq} twice to get
\begin{gather}
\label{eq:dN_path_int_1}
    e^{-\frac{\eps_2}{2}\dN}\dN' = C\frac{\Pphi(N)}{\epseff_1(N)} = C\PR(N) \\
\label{eq:dN_path_int_2}
    \Rightarrow X(k') = \int_{\kini}^{k'} e^{-\frac{\eps_2}{2}\dN}\frac{\dd \dN}{\dd \ln k''} \dd \ln k'' = C\sig{k'}^2 \, ,
\end{gather}
where I followed the conventions for $\epseff_1(N)$, $X$, $\sig{k'}$, and $\dd N = \dd \ln k$ from above. Setting the constant of integration $C=X(k)/\sig{k}^2$ to fix the final value $X(k)$ (and thus $\dN$) at $k'=k$, this is the exact result \eqref{eq:p_X_cond_2} for the conditional expectation value $\overbar{X}(k')_{|k}$: the method of \cite{Tomberg:2022mkt} correctly produces the most probable $X$-paths in constant roll. However, the mean $\dN$ paths slightly differ from these due to the non-linear relationship between $X$ and $\dN$, see \eqref{eq:dN_bar_cond}; taking the result \eqref{eq:dN_path_int_2} and converting to $\dN$ by inverting \eqref{eq:X} only reproduces \eqref{eq:dN_bar_cond} in the $\sigC{k'}{k} \to 0$ limit. In \eqref{eq:xi_distr}, the difference can be attributed to the different volumes in the $\xiN$-space one has to integrate over to catch all the paths in an $X$ bin versus a $\dN$ bin.

\section{Modified Higgs models} 
\vspace{-0.5cm}
\label{sec:higgs_models}
In \cite{Figueroa:2020jkf, Figueroa:2021zah}, three inflection point models of modified Higgs inflation were studied. A supercomputer was used to numerically solve the probability distributions $p[\dN(k)]$ at a scale $k=k_\text{PBH}$ somewhat past the power spectrum peak, from the full equations~\eqref{eq:bg_eom}, together with simultaneously evolved short-wavelength perturbations to provide the stochastic noise. All the example models follow the dual ultra-slow-roll--constant-roll behavior, with power spectra similar to Figure~\ref{fig:PR}. Starting from the power spectra, I have re-estimated $p[\dN(k)]$ and the PBH abundance by the methods of this paper and compared them to the numerical results\footnote{I thank Sami Raatikainen for providing the data of \cite{Figueroa:2021zah} in its original accuracy.} of \cite{Figueroa:2020jkf, Figueroa:2021zah}. I present the results in Figure~\ref{fig:p3}. The matches are excellent.

To estimate the PBH abundance, \cite{Figueroa:2021zah} extrapolated $p[\dN(k)]$ beyond the collapse threshold with an exponential fit $e^{A-BN}$. The fit soon starts to deviate from the curve \eqref{eq:p_Delta_N} which turns further down. As a consequence, the fit overestimates the PBH abundance in all but the solar mass case, where the tail was resolved reliably beyond the collapse threshold. Table~\ref{tab:higgs_models} lists key figures for the three models.

\bibliographystyle{apsrev4-1}
\bibliography{stoc}

\end{document}